\documentclass[conference]{IEEEtran}
\IEEEoverridecommandlockouts

\usepackage{cite}
\usepackage{amsmath,amssymb,amsfonts}
\usepackage{algorithmic}
\usepackage{graphicx}
\usepackage{textcomp}
\usepackage{xcolor}
\usepackage{booktabs}
\usepackage{url}
\def\BibTeX{{\rm B\kern-.05em{\sc i\kern-.025em b}\kern-.08em
    T\kern-.1667em\lower.7ex\hbox{E}\kern-.125emX}}

\newcommand{\sysname}{\textsc{Anon}}
\newcommand{\crateVer}{0.10.0}
\newcommand{\nTests}{241}
\newcommand{\nDefectClasses}{24}
\newcommand{\nFamilies}{seven}
\newcommand{\nSystemicClasses}{eight}
\newcommand{\nExternalFrameworks}{three}
\newcommand{\nFamiliesCross}{two}
\newcommand{\nRevertVerified}{two}
\newcommand{\nFlakeRuns}{10}
\newcommand{\nSuiteReruns}{three}
\newcommand{\otpTag}{OTP-28.0}
\newcommand{\otpDocsVer}{28.5.0.4}
\newcommand{\ractorRev}{1af0aaad}
\newcommand{\actixVer}{0.13.5}
\newcommand{\mailboxDefault}{64}

\renewenvironment{abstract}{%
  \normalfont\small
  \noindent\textbf{Abstract}\par\smallskip
  \noindent\ignorespaces
}{\par}

\begin{document}

\title{Supervision Without Blocking:\\An Evidence-Driven Redesign of OTP-Style Fault Tolerance for Cooperative Runtimes}

\author{\IEEEauthorblockN{Sam Uwejan}
\IEEEauthorblockA{M.S. in Artificial Intelligence Program\\
Department of Computer Science\\
Angelo State University\\
San Angelo, TX, USA\\
suwejan@angelo.edu}
\and
\IEEEauthorblockN{Erdogan Dogdu}
\IEEEauthorblockA{Department of Computer Science\\
Angelo State University\\
San Angelo, TX, USA\\
erdogan.dogdu@angelo.edu}}

\maketitle

\begin{abstract}
Actor frameworks on mainstream runtimes advertise Erlang/OTP-style supervision, but OTP's consistency contracts assume a synchronous, single-threaded supervisor, and ports inherit that assumption's absence as unexamined risk. We evaluated every supervision guarantee against pinned OTP documentation and source read verbatim, against the source of \nExternalFrameworks{} further Rust actor frameworks at fixed revisions, and against the actor-model literature, recording the evidence behind each design decision. Blind adversarial audits and the cross-framework study surfaced \nDefectClasses{} defect classes, of which \nSystemicClasses{} reduce to one root cause: reconstructing, by explicit mechanism, the atomicity a synchronous supervisor gets implicitly. We introduce a supervision runtime that closes all \nDefectClasses{} cataloged classes through supervisor-owned completion values, per-incarnation identity, and a single lifecycle state machine, validated by \nTests{} tests including reverted-fix reproductions. The result holds OTP's verbatim-verified contracts while adding two properties OTP does not offer: a supervisor that remains responsive throughout group restarts, and message handlers that are never cancelled mid-execution. We delimit the forcing guarantees no cooperative runtime can honor.
\end{abstract}

\section{Introduction}

Fault tolerance in the Erlang/OTP tradition rests on supervision trees \cite{armstrong2003reliable}. A supervisor process owns the lifecycle of its children and restarts them according to a declared policy. This design has migrated far beyond the BEAM virtual machine. Actor libraries for Rust, the JVM, and .NET advertise supervision as a headline feature, and production systems increasingly depend on those implementations honoring the contracts the original defines.

Existing implementations inherit OTP's semantics wholesale or partially, without evaluating which guarantees deserve keeping and which reflect accidents of the BEAM's architecture. An OTP supervisor is a single-threaded process that blocks for the duration of each restart and termination step it performs, and its consistency contracts are silent consequences of that blocking. Frameworks on cooperative schedulers such as Tokio, an asynchronous Rust runtime \cite{tokio}, drop the blocking to keep their supervisors responsive. None of the implementations we examined priced what that removal costs or what it makes possible.

We took the opposite route and treated every supervision guarantee as a design decision that must earn its place on evidence. We read the governing OTP documentation and supervisor source verbatim at pinned versions. We examined the source of \nExternalFrameworks{} further Rust actor frameworks at fixed revisions, drew on the actor-model literature, and subjected our own framework to blind adversarial audits, recording the evidence behind every decision. The audits and the cross-framework study surfaced \nDefectClasses{} defect classes and one dominant root cause. All four codebases, at the revisions we examined, had independently adopted push-based lifecycle signaling, whose losses and reorderings are the unpaid price of an asynchronous supervisor.

The resulting runtime, \sysname{}, pays that price deliberately. The supervisor holds each child's completion as a value returned by a task set it owns. Every instance carries an identity drawn from a monotonic counter. All liveness bookkeeping collapses into one state machine with a single transition function. Teardown reproduces the documented OTP asymmetry by which a kill delivered over a link arrives trappable while a direct kill does not \cite{otp28processes,otp28erlangmodule}. Where the evidence favored departing from OTP, we departed. The supervisor never blocks, and a message handler, once begun, is never cancelled.

The outcome closes all \nDefectClasses{} cataloged classes, validated by \nTests{} tests that include regression reproductions failing with the fixes reverted. The runtime holds OTP's verbatim-verified contracts alongside two properties OTP does not offer, namely full supervisor responsiveness during group restarts and handler-turn isolation. The guarantee analysis states each kept, added, and traded property with its evidence and its price.

\begin{itemize}
\item \textbf{An evidence-driven design method.} We evaluate every supervision guarantee against pinned primary sources read verbatim, the source of \nExternalFrameworks{} competing frameworks at fixed revisions, and blind adversarial audits, and we record the evidence behind each decision, including the decisions to depart from OTP.
\item \textbf{A supervision race taxonomy.} We catalog \nDefectClasses{} defect classes reducing to \nFamilies{} mechanism families with one dominant root cause, and we document instances of \nFamiliesCross{} families recurring in the examined external frameworks.
\item \textbf{An ownership-based supervision runtime.} We introduce \sysname{}, in which completions are supervisor-owned values, identity is per-incarnation, liveness is a single state machine, and termination follows a signal-first cascade, closing every cataloged class while the supervisor stays responsive.
\item \textbf{A three-way guarantee analysis.} We state which OTP contracts the runtime reproduces with verbatim-source verification, which properties it provides beyond OTP, and which forcing guarantees no cooperative runtime can honor, each with its evidence and its cost.
\end{itemize}

\section{Related Work}

\textbf{The actor model.} Hewitt et al. introduced actors as general primitives of concurrent computation \cite{hewitt1973actor}, and Agha developed the model into a semantics of isolated state, asynchronous messages, and dynamic topology \cite{agha1986actors,agha1997foundation}. We claim no novelty in the actor model itself; every primitive our runtime exposes descends from this lineage. Our delta is an account of what the model's fault-tolerance layer demands from a host scheduler.

\textbf{Supervision in Erlang/OTP.} Armstrong framed reliability as isolation plus supervised restart \cite{armstrong2003reliable}, and the OTP supervisor behaviour implements this contract as a synchronous, single-threaded process \cite{otp28supprinc,otp28supervisor}. Prior descriptions of OTP treat the supervisor's synchronicity as an implementation detail. Our delta is to show it is load-bearing: we trace \nSystemicClasses{} defect classes in ports directly to its absence.

\textbf{Actor frameworks outside the BEAM.} Orleans reframed actors as virtual, automatically activated entities \cite{bykov2011orleans}, and Pony co-designed its type system and runtime around capability-safe, garbage-collected actors \cite{clebsch2015deny,clebsch2017orca}. In the Rust ecosystem, actix supervises exactly one actor by polling it inline \cite{actix}, ractor delegates restart policy entirely to the implementor \cite{ractor}, and the archived bastion attempted full supervision trees over push-based signaling \cite{bastion}. Our delta is a source-level comparison of the three Rust frameworks' lifecycle signaling at pinned revisions, showing convergence on a shape none of them chose deliberately.

\textbf{Studies of actor concurrency defects.} Bagherzadeh et al. mined and classified real-world actor bugs across several ecosystems \cite{bagherzadeh2020actor}. Their taxonomy classifies application-level defects observed in user code. Our delta is a taxonomy one level down: defects in the supervision runtime itself, derived from audits and cross-framework source examination rather than issue mining.

\textbf{Structured concurrency.} The structured-concurrency argument holds that a task's lifetime should be bounded by a scope its parent owns \cite{smith2018structured}, a discipline Tokio exposes through owned task sets \cite{tokio}. We borrow this discipline rather than invent it. Our delta is applying it to supervision bookkeeping itself: completions become values owned by the supervisor, which removes the lost-signal class by construction.

\section{Study Method and Race Taxonomy}
\label{sec:method}

\subsection{Audit Protocol}

The taxonomy rests on three evidence sources. Each was designed to avoid anchoring on what the framework's documentation claims.

\textbf{Blind adversarial audits.} Independent reviews examined the framework's full source with a correctness-only brief and no access to its design documents. Each review was required to cite an exact file and line and a concrete failure interleaving for every reported defect. We confirmed every accepted finding by re-tracing the interleaving in the cited code before counting it. Reviews ran in independent pairs with complementary scopes, one over the whole codebase and one over the change under review, so that a defect missed by one scope could still surface in the other.

\textbf{Primary-source semantics verification.} Every claim the design makes about OTP behavior was verified against pinned primary sources rather than secondary descriptions. The sources are the Erlang reference manual and erlang module documentation at version \otpDocsVer{} \cite{otp28processes,otp28erlangmodule}, and supervisor.erl at tag \otpTag{} with line-anchored citations \cite{otp28supervisor}. Three design-critical behaviors were confirmed verbatim this way, namely the supervisor's unlink-and-flush discipline when terminating a child, the escalation of a shutdown timeout to a direct untrappable kill, and the re-inclusion of previously terminated children in group restarts.

\textbf{Cross-framework source examination.} We examined \nExternalFrameworks{} Rust actor frameworks at pinned revisions, namely ractor at commit \ractorRev{} \cite{ractor}, actix \actixVer{} \cite{actix}, and the unmaintained bastion at its final state \cite{bastion}. Each examination answered fixed questions about restart synchronicity, death signaling, kill delivery, and teardown on abort, citing file and line in the examined revision.

\subsection{The Root Cause}
\label{sec:rootcause}

An OTP supervisor is one sequential process \cite{otp28supprinc}. When a child dies, the supervisor handles the exit message to completion before it dequeues anything else, so a manual command observes either a fully processed death or none at all. The one deliberate exception proves the sensitivity of this discipline. A failed restart attempt requeues its retry through the supervisor's own mailbox so that other requests can interleave before the retry \cite{otp28supervisor}, and OTP accepts the resulting window rather than block indefinitely. A supervisor on a cooperative runtime that refuses to block must interleave death processing, manual commands, and restart completions everywhere, and every piece of shared state those steps touch becomes a race surface. Of the \nDefectClasses{} classes we cataloged, \nSystemicClasses{} reduce to exactly this reconstruction burden.

\subsection{Mechanism Families}
\label{sec:taxonomy}

The \nDefectClasses{} classes reduce to \nFamilies{} families, ordered by the mechanism that fails.

\textbf{Stale liveness classification.} A manual stop issued while the same child's crash notification is still queued reads a stale liveness flag and reclassifies the crash as a manual stop. In the studied framework this bypassed the restart budget for a crash-looping child. In one configuration it permanently deleted a restartable worker's specification.

\textbf{Lossy completion signaling.} Death and restart notifications sent over bounded channels can be dropped at capacity or deferred behind unrelated work. The studied framework's watcher acknowledgments and restart completions both carried this exposure. Among the examined frameworks, bastion discards lifecycle send results outright on its restart path, while ractor and actix avoid this specific exposure only by making the equivalent step inline in the supervisor, at the cost of blocking it.

\textbf{Termination-order violations.} Observers could see a child's terminal state before its registry name was released. A restart racing that window failed on a name collision, and a caller acting on a completed termination could still be refused.

\textbf{Incarnation identity confusion.} Fresh spawns reused a constant incarnation identity. A straggling death report from a previous instance could therefore mark its live successor dead once termination became fast enough to expose the window.

\textbf{Group-restart interleavings.} Group restarts interleaved with solo restarts and manual stops in ways a sequential supervisor cannot express. A restart attempt failing during an unrelated group stop wedged the group permanently, and a manual termination during a group stop violated the caller's completion contract.

\textbf{Teardown orphaning.} A force-aborted supervisor never ran its own teardown, orphaning its subtree. The examined frameworks carried instances of the same family, namely ractor's abort path before its 2026 lifecycle-guard fix and bastion's task handles, which detach rather than cancel on drop. Auxiliary forwarder tasks for timers and streams also outlived their registrations or died silently on panic.

\textbf{Escalation capacity coupling.} Stop and kill signals traveled paths that could await mailbox capacity, so the staged escalation intended to bound a shutdown could itself block without a deadline.

\section{An Ownership-Based Supervision Runtime}
\label{sec:design}

The redesign replaces notification with ownership across every lifecycle mechanism. This section presents the six mechanisms and maps each to the families of Section~\ref{sec:taxonomy}.

\subsection{A Monotone Stop Lane}

Each actor owns a stop lane, a single-value watch cell holding the highest-severity pending stop request. A generation counter increments on every accepted raise. Raising the lane is synchronous and infallible, severity never decreases, and the run loop observes the lane with the highest priority in its event selection. This removes the escalation capacity coupling family. No stop or kill signal can await mailbox capacity, which mirrors the OTP treatment of exit signals as a class apart from ordinary messages \cite{otp28processes}. Signals are still delivered only at turn boundaries, so an in-flight message handler is never cancelled mid-execution. Section~\ref{sec:fidelity} develops this deliberate divergence from BEAM preemption.

\subsection{Owned Death and Restart Completions}

The supervisor owns a task set of watcher tasks, per-child monitors that observe a child task's completion, one per child incarnation. An incarnation is a monotonically numbered instance of a child identity, minted at each spawn or restart. Each watcher writes the child's real termination reason into a per-incarnation fate cell, a write-once slot holding that instance's terminal outcome, after the child's task fully completes. Restart attempts run in a second supervisor-owned task set whose results are returned values, not messages. Both push-notification paths of the original design are deleted rather than hardened, which removes the lossy completion signaling family by construction. A completion the supervisor owns cannot be dropped at capacity, and an attempt abandoned for any reason resolves through the drop of its owned handle. Because the fate cell is written only after the child's task has run all of its destructors, a recorded fate also proves the child's registry name is free, which removes the termination-order family.

\subsection{A Single Lifecycle Ledger}

All per-child liveness bookkeeping collapses into one state machine, the ledger, over four states, namely running, stopping, restarting, and down. One transition function rejects every undeclared transition, and the scattered boolean flags whose staleness produced the misclassification family are deleted. Manual stops commit synchronously, with zero suspension points between recording the intent and raising the signal. Classification of the completed stop uses the fate cell's real termination reason. A crash that races a manual termination is absorbed into it, matching the supervisor's verbatim unlink-and-flush discipline \cite{otp28supervisor}. A crash racing a restart-eligible stop is charged to the restart budget.

\subsection{Per-Incarnation Identity}

Every registration, fresh spawns included, draws its incarnation from one monotonic counter. Every fate, watcher, and adoption decision is keyed by the pair of child identity and incarnation, so a straggling death report from a superseded instance no longer matches its successor. This removes the identity confusion family. The BEAM enforces the analogous property at the virtual-machine level, stating that a process identifier cannot be released for reuse until everything regarding the process has been released \cite{otp28processes}.

\subsection{A Signal-First Termination Cascade}

Teardown semantics follow the OTP documentation read verbatim. A direct kill is untrappable and immediate, and the runtime rewrites the dying process's exit reason from kill to killed before it reaches the process's links, so linked processes receive a trappable killed signal rather than the untrappable kill \cite{otp28erlangmodule}. The redesign reproduces this asymmetry structurally. A killed supervisor does not walk its subtree. It drops its supervision state, and each child's guard, a destructor-driven handle firing when the child's supervision state is dropped, raises kill on that child's stop lane. A cooperative child observes the signal at its next turn boundary and runs its own teardown one level down. A child that fails to exit within a bounded grace window is force-aborted by the per-system reaper, a single background task that aborts children exceeding their grace deadline. Escalated kills arising from an exhausted shutdown timeout take the same path, matching the supervisor source, which escalates a shutdown timeout to a direct untrappable kill \cite{otp28supervisor}. This removes the teardown orphaning family, together with a guard rule that every per-incarnation resource is reconstructed when a restarted instance is adopted.

\subsection{Group Restarts over the Ledger}

Group restart phases are expressed against the ledger rather than against ad hoc flags. A member whose restart attempt fails during an unrelated group stop is settled to the down state, from which the group's restart chain legally revives it. A previously terminated member re-enters the chain in its original slot order, matching the supervisor source's treatment of terminated children in group restarts \cite{otp28supervisor}. A manual termination during a group stop settles the ledger synchronously, so the caller's completion contract holds even within a single message handler. These rules remove the group interleaving family.

\section{Guarantee Analysis}
\label{sec:fidelity}

A runtime that claims OTP-style supervision owes its users a precise account of its guarantees. We state them in three categories, namely the OTP contracts the runtime reproduces with verbatim-source verification, the properties it provides that OTP does not, and the guarantees no cooperative runtime can honor. Table~\ref{tab:fidelity} summarizes the analysis, and the text states the reasoning for every entry involving a decision.

\begin{table}[t]
\caption{OTP process and supervision guarantees relevant to child lifecycle management, not an exhaustive account of OTP semantics; status reflects the redesigned runtime. Verbatim-source entries cite the pinned document stating the guarantee.}
\label{tab:fidelity}
\centering
\small
\begin{tabular}{p{0.46\columnwidth}p{0.42\columnwidth}}
\toprule
Guarantee & Status \\
\midrule
Exit signals outrank ordinary messages \cite{otp28processes} & Reproduced: stop lane at highest selection priority \\
Direct kill untrappable; linked processes receive trappable killed \cite{otp28erlangmodule} & Reproduced structurally: cascade inversion plus signal-first guards \\
Shutdown timeout escalates to direct kill \cite{otp28supervisor} & Reproduced on the same path as configured kills \\
Termination call returns with the child fully gone \cite{otp28supervisor} & Reproduced: fate cell implies name release \\
Terminated members rejoin group restarts \cite{otp28supervisor} & Reproduced in slot order \\
Restart intensity accounting \cite{otp28supprinc} & Reproduced; crash-race reclassification closed \\
Supervisor responsive throughout group restarts & Provided beyond OTP, whose supervisor blocks per attempt \\
Handler-turn isolation: a begun handler always completes & Provided beyond OTP, which cancels mid-reduction \\
Completion contracts held without blocking the supervisor & Provided beyond OTP: termination truth plus responsiveness \\
Preemptive kill of a running handler & Not reproduced: turn-boundary delivery by decision \\
Kill overrides a non-yielding computation & Not reproducible: cooperative scheduling bounds all forcing \\
Process death survives host-process kill & Out of scope for any in-process runtime \\
\bottomrule
\end{tabular}
\end{table}

\textbf{Reproduced contracts.} Each entry in the first group was verified against the pinned primary source before the mechanism reproducing it was designed, and Section~\ref{sec:design} names the mechanism per contract. The termination truth contract deserves emphasis. When a termination call returns, the child is fully gone and its name is free, so a delete and same-name respawn succeed immediately, even inside the same message handler. OTP provides this contract by blocking its supervisor for the duration of the termination. The redesign provides it while the supervisor keeps processing every other event.

\textbf{Properties beyond OTP.} OTP buys consistency within each restart attempt by blocking the supervisor for that attempt's duration, and a failed attempt requeues its retry so that other messages can interleave before it \cite{otp28supervisor}. The redesigned supervisor never blocks at all. It answers status queries and manual commands throughout group restarts, terminations, and escalations. Section~\ref{sec:rootcause} states the price, since \nSystemicClasses{} defect classes exist only because the blocking supervisor's implicit atomicity must be reconstructed by explicit mechanism. The taxonomy and the ledger are that reconstruction, and we report the property as a paid-for trade rather than a free improvement. The second added property is handler-turn isolation. The BEAM delivers an untrappable kill preemptively, without waiting for the target to reach a user-code safe point, and one examined framework imitates this by racing a cancellation against the handler's next suspension point \cite{ractor}. We decline the imitation on recorded evidence. Cancelling a handler mid-execution drops its local state at an arbitrary await point and exports a cancellation-safety obligation into every user handler. The redesign therefore guarantees the stronger per-handler property that a begun handler always completes, at the cost of kill latency bounded by one handler turn rather than one scheduling step.

\textbf{The forcing boundary.} Erlang's kill is guaranteed by a virtual machine the language owns. An in-process runtime on a host language owns neither the scheduler nor the process. Its strongest primitive is task abort at a yield point, a handler that never yields defeats both the signal and the abort backstop, and an operating-system kill of the host runs no teardown at all. Every stronger claim in any implementation's documentation overstates what the substrate permits.

\section{Evaluation}
\label{sec:evaluation}

We ask whether the redesign closes the cataloged defect classes in the studied framework. We answer with executable evidence and state the boundary of that evidence directly.

\subsection{Setup}

All evidence comes from the framework's test suite at version \crateVer{}, running on Tokio \cite{tokio} under its multi-threaded scheduler. The suite comprises \nTests{} passing tests across unit, integration, and documentation tests. Race regression tests were additionally run \nFlakeRuns{} consecutive times each to expose scheduling-dependent flakiness, and the full suite was run \nSuiteReruns{} consecutive times per change batch.

\subsection{Per-Family Regression Evidence}

Each mechanism family of Section~\ref{sec:taxonomy} is covered by integration tests that drive the real failure path rather than asserting on types or constructors. The misclassification family is exercised by racing a genuine crash against a manual stop in both intent directions and asserting the budget charge. The signaling family is exercised by mass simultaneous deaths far exceeding any prior channel capacity, with zero lost fates. The ordering family is exercised by a same-handler terminate, delete, and same-name respawn sequence asserted inside one message dispatch. The identity family is exercised by that same sequence looped under a deliberately delayed straggler fate from the previous incarnation. The group family is exercised by the wedge interleaving, the manual-override contract, and slot-order revival of a terminated member. The orphaning family is exercised by killing supervisors mid-restart across every attempt phase and asserting name release, and by cascade tests asserting that no descendant remains observably running. The escalation family is exercised by kill delivery under a mailbox saturated at its default capacity of \mailboxDefault{} messages.

\nRevertVerified{} classes carry the strongest form of evidence. Their regression tests were verified to fail against the pre-fix code by temporarily reverting the fix, then re-applied and observed to pass across \nFlakeRuns{} runs. The remaining classes are closed by construction, in the sense that the mechanism they exploited no longer exists, and their tests pin the replacement behavior.

\subsection{What This Evaluation Does Not Show}

The evaluation demonstrates correctness properties, not performance. Throughput, latency, and memory comparisons against other frameworks are a recorded follow-up, not a claim of this paper, and no number in this paper is a performance number. Test-based evidence also cannot demonstrate the absence of further defect classes, a limit Section~\ref{sec:threats} develops.

\section{Threats to Validity}
\label{sec:threats}

\textbf{Single primary subject.} The taxonomy's depth comes from one framework we control, and the cross-framework evidence covers \nExternalFrameworks{} further codebases at single pinned revisions. Classes that require mechanisms none of these four frameworks implement would not surface, so the taxonomy is a lower bound, not a census.

\textbf{Evidence standard.} All closure claims rest on tests, including adversarially constructed interleavings, but tests cannot prove the absence of unexplored schedules. We mitigated this with repeated runs, revert verification for two classes, and audits that were blind to the design documents, and we scope every closure claim to the cataloged classes.

\textbf{Audit reproducibility.} The adversarial audits followed a fixed brief and citation discipline, but independent auditors examining the same code could catalog a different class set or draw family boundaries differently. The per-class code citations and failure interleavings are recorded, so the findings are checkable even where the categorization is judgment.

\textbf{Author position.} We audit and redesign our own framework, which risks favorable framing. The mitigations are structural: findings were only counted with a confirmed code-level trace, the cross-framework examinations quote the examined revisions, and the fidelity analysis names what the redesign does not achieve with the same prominence as what it does.

\section{Discussion and Limitations}
\label{sec:discussion}

\textbf{Why the convergence happened.} Four independently developed frameworks arrived at push-based lifecycle signaling, and we attribute this convergence to shared constraints of the async substrate rather than to coincidence. Channels are the idiomatic communication primitive on an async runtime, so announcing a death over a channel requires the least additional code. The ownership alternative requires the supervisor to hold and poll task sets, which the ecosystem's structured-concurrency primitives made ergonomic only recently \cite{tokio,smith2018structured}. The taxonomy indicates that the idiomatic choice and the correct choice diverge exactly where lifecycle events carry obligations rather than data.

\textbf{Generality of the taxonomy.} The mechanism families are stated in framework-neutral terms, and the cross-framework examination found instances of \nFamiliesCross{} families outside our own code, namely lossy completion signaling and teardown orphaning. We expect the families to apply to any never-blocking supervisor over push signaling on any cooperative runtime, but our evidence covers only four Rust codebases.

\textbf{Limitations.} The redesign's guarantees are scoped by the guarantee analysis. Turn-boundary delivery bounds kill latency by one handler turn, a never-yielding handler defeats all forcing, and a host-process kill runs no teardown. The evaluation is correctness-only, so the runtime cost of ownership bookkeeping relative to push signaling is unmeasured. Peer-to-peer monitoring between actors, as opposed to vertical supervision, is not yet designed. Extending the ownership discipline to an observation plane in which no observer may extend any actor's lifetime is open.

\section{Conclusion}
\label{sec:conclusion}

OTP's supervision guarantees depend on a synchronous supervisor, and implementations on cooperative runtimes must reconstruct that atomicity themselves through explicit mechanism. We treated every guarantee as a design decision to be earned on evidence from pinned primary sources, competing framework source, and blind adversarial audits. We cataloged \nDefectClasses{} defect classes from those audits and \nExternalFrameworks{} source examinations, and reduced them to \nFamilies{} mechanism families with one dominant root cause. We closed every cataloged class in the studied framework by replacing push signaling with supervisor-owned completions, per-incarnation identity, one lifecycle state machine, and a signal-first cascade derived verbatim from pinned OTP sources. The result is validated by \nTests{} tests, including \nRevertVerified{} revert-verified reproductions, and holds OTP's verbatim-verified contracts alongside supervisor responsiveness during group restarts and handler-turn isolation, two properties OTP does not offer. The guarantee analysis states each kept, added, and traded property at its earned strength. Performance measurement and a peer-observation plane under the same ownership discipline remain recorded follow-ups.

\bibliographystyle{IEEEtran}
\bibliography{refs}

\end{document}